\documentclass[
 aip,
 amsmath,amssymb,
 reprint,
]{revtex4-1}

\usepackage{graphicx}
\usepackage{dcolumn}
\usepackage{bm}
\usepackage{titlesec}
\usepackage[utf8]{inputenc}
\usepackage[T1]{fontenc}
\usepackage{mathptmx}
\usepackage{etoolbox}
\usepackage{xcolor}
\usepackage{graphicx}
\usepackage[percent]{overpic}
\usepackage{subcaption}
\usepackage{nicefrac}
\makeatletter
\def\@email#1#2{
 \endgroup
 \patchcmd{\titleblock@produce}
  {\frontmatter@RRAPformat}
  {\frontmatter@RRAPformat{\produce@RRAP{*#1\href{mailto:#2}{#2}}}\frontmatter@RRAPformat}
  {}{}
}
\makeatother
\begin{document}

\preprint{AIP/123-QED}

\title{Collision Cascade Morphology and Defect Nucleation in Pure Tungsten and an Equimolar Refractory W-Ta-Cr-V Alloy}
\author{M. Fedorov}
\author{F. J. Dominguez-Gutierrez}
\author{D. Kalita}
\author{L. Kurpaska}
\affiliation{ 
National Centre for Nuclear Research, NOMATEN CoE, ul. Andrzeja Soltana 7, 05-400 Świerk, Poland
}

\date{\today}
\begin{abstract}
Collision cascades and defect-formation mechanisms in pure W and the WTaCrV refractory high-entropy alloy were investigated using molecular dynamics simulations over primary knock-on atom (PKA) energies of 0.1--100~keV. Machine-learning tabulated Gaussian approximation potentials (tabGAP), complemented by short-range Ziegler-Biersack-Littmark (ZBL) interactions, were employed. Analysis of 
the heat spike lifetime, affected region, atomic relocations and surviving defects reveals an importance of subcascade formation and presence of light elements in the alloy.
Cascade splitting into subcascades contributes to the distribution of the PKA energy over a larger region in the material, nucleating less dislocations and smaller voids -- latter being present only in W and absent in WTaCrV.
Since cascade splitting is more frequently observed in WTaCrV, the alloy shows better resistance to primary irradiation damage compared to pure W.
Apart from cascade splitting, analysis of interstitials shows that Cr and to lesser extent V, being lighter and smaller elements, accommodate the damage in form of point defects, contributing to the mitigation of formation of the dislocation loops.

\end{abstract}

\maketitle

\section{\label{sec:level1} Introduction}

High entropy alloys (HEAs)\cite{Cantor2004,Yeh2004}, including W-based 
refractory high entropy alloys have been of interest to the fusion 
community as potential plasma-facing materials in the first wall and 
the divertor of the future fusion 
reactors\cite{li2021tungsten,hatler2025currentopinioninsolidstateandmaterialsscience,SUN2025121435}.
As is the general appeal of HEAs, W-based HEAs have been reported to 
have excellent mechanical properties at high temperatures and 
irradiation-resistance properties \cite{wang2021tungsten,yoo2024intjminermetallmater,elatwani2019,korpe2025}, as well as great corrosion-resistance properties \cite{zhang2022surfaceandcoatingstechnology} -- all important 
for materials designed to work in harsh environments. The 
W$_{38}$Ta$_{36}$Cr$_{15}$V$_{11}$ alloy has shown a lack of 
dislocation loops after irradiation up to 8~dpa and a slight 
irradiation hardening \cite{elatwani2019}. The original second phase 
Cr- and V-rich precipitates grew under the influence of irradiation.
Experimental observations of He-ion irradiated non-equimolar W$_{15}$Ta$_{15}$Cr$_{35}$V$_{35}$ also show reduced He bubble formation and growth, attributed to a combination of modified diffusion due to lattice distortion and the formation of carbon-vacancy complexes that inhibit He aggregation inside vacancies\cite{wang2025actamaterialia,KALITA2025113751}.
Amorphous films of WTaCrV can be produced and also have irradiation-resistance properties \cite{TUNES2023101796}.
The equiatomic WTaCrV alloy has been shown experimentally to have 
higher resistance to surface damage under He-ion irradiation compared 
to pure W\cite{ELATWANI2021100599}.
Theoretical investigations confirm lower tendency to fuzz formation, He 
bubble formation and He bubble growth under irradiation due to reduced formation and binding energies of vacancy-helium clusters, as well as 
reduced thermodynamic driving force and kinetically suppressed He 
motion in WTaCrV and WTaCrVTi HEAs compared to pure W \cite{xiong2024journalofnuclearmaterials,shi2025physscr,shi2026computationalmaterialsscience}.

Theoretical investigation with phase-field distortion dynamic model suggested introduction of low local unstable stacking fault energies as being behind the mechanism for reduced irradiation hardening in WTaCrV alloy\cite{li2025internationaljournalofplasticity}.
High-throughput experimental exploration of W-based HEAs has shown that 
compositional complexity reduces the nanofuzz 
grow\cite{yan2026journalofnuclearmaterials}. It has also shown that addition of Ta enhances the resistance to He-ion irradiation, whereas addition of V reduces the resistance.
Molecular dynamics simulations of the interaction of collision cascades with pre-existing dislocations in WTaCrV show that the presence of latter can reduce the effects of irradiation\cite{li2024japplphys}, which suggests that mechanical treatment can further improve irradiation resistance of W-based HEAs.
Semi-empirical Finnis-Sinclair interatomic potential was developed and tested for W$_{38}$Ta$_{36}$Cr$_{15}$V$_{11}$ alloy, showing fewer dislocation loops formation and smaller interstitial clusters in the alloy compared to pure W\cite{chen2023journalofnuclearmaterials}.

In this work, we perform molecular dynamics simulations with a newly-developed machine-learned interatomic potential\cite{byggmästar2026arxiv} to systematically investigate the energy-dependent evolution of collision cascades and primary defect formation over a 0.1--100~keV range of primary knock-on atom (PKA) energies in pure W and the equiatomic WTaCrV alloy, which has been shown experimentally to have high resistance against defect accumulation \cite{KALITA2023101513,KALITA2025113751}.
Particular emphasis is placed on establishing the connection between PKA energy, cascade morphology, and the resulting defect structures.
The duration and geometric properties of the thermal spike, cascade-induced atomic replacements, as well as surviving 1D, 2D and 3D defects population and elemental composition are quantified, and dependencies between them are analyzed, providing insight for the mechanisms that may contribute to the enhanced irradiation tolerance of refractory multi-component alloys.
Through this combined temporal, structural, and energetic analysis, we elucidate how chemical complexity modifies energy dissipation and defect organization in WTaCrV relative to elemental W, providing an atomistic framework for understanding the mechanisms that may contribute to the enhanced irradiation tolerance of refractory multi-component alloys.

\section{Computational method}
\label{sec:methods}

In this work, interatomic interactions are described using a recently developed machine-learning interatomic potential based on a tabulated implementation of a low-dimensional Gaussian Approximation Potential (tabGAP) framework \cite{PhysRevMaterials.6.083801, PhysRevB.104.104101, BYGGMASTAR2025121276,byggmästar2026arxiv}.
Within the GAP formalism, the total energy of the system is expressed as a sum of local atomic energy contributions, each dependent on the surrounding atomic environment and trained against reference data obtained from density functional theory (DFT) calculations \cite{PhysRevMaterials.6.083801, BYGGMASTAR2025121276}.
Unlike high-dimensional descriptor approaches such as the Smooth Overlap of Atomic Positions (SOAP) formalism, the low-dimensional GAP model employed here represents atomic environments using a compact set of physically motivated descriptors.
These include pairwise distance terms, angular three-body correlations, and a scalar many-body density contribution analogous to that used in the embedded-atom method \cite{PhysRevB.104.104101,PhysRevMaterials.7.043603}.
This reduced descriptor set captures the essential physics of metallic bonding, namely coordination dependence, angular interactions, and local electron density effects, while enabling an efficient and compact representation of the potential energy surface.
The tabGAP approach leverages this low-dimensional representation by pre-tabulating the descriptor-dependent energy contributions on regular grids, which are evaluated during molecular dynamics simulations using spline interpolation \cite{PhysRevB.104.104101}.
Moreover, the tabulated formulation ensures smooth force fields and enhanced numerical stability, which are essential for large-scale and long-time molecular dynamics simulations, particularly those involving high-energy collision cascades and defect evolution.

At the short interatomic distances encountered during high-energy collision cascades, interatomic interactions are dominated by strong screened Coulomb repulsion between atomic nuclei.
This regime is well described by the universal Ziegler–Biersack–Littmark (ZBL) potential, which provides an accurate representation of nuclear stopping through a screened pairwise interaction that is independent of chemical bonding effects \cite{Ziegler1985}.
The ZBL potential is therefore widely employed in MD simulations of irradiation damage to correctly capture close-range repulsive interactions during ballistic collisions, where conventional interatomic potentials are no longer valid \cite{SIMEONE2026122514}.
In the present work, the ZBL potential is smoothly joined to the machine-learning interatomic potential at short interatomic separations, ensuring a physically consistent description of interatomic forces across the full range of relevant distances. 

MD simulations were performed using the Large-scale Atomic/Molecular Massively Parallel Simulator (LAMMPS) \cite{LAMMPS}.
For collision cascade simulations, the simulation cell must be sufficiently large to accommodate the trajectory of energetic recoiling atoms in the keV energy range
\cite{Nordlund2018,USTRZYCKA2024104118,1kyk-k6n5,SIMEONE2026122514}. The dimensions of the simulation cells employed for pure W at each PKA energy are summarized in Table~\ref{table_MD}.
The initial (001) W samples were constructed with a lattice parameter of 0.316~nm.
Prior to the cascade simulations, all structures were subjected to energy minimization to obtain relaxed
configurations.
The minimization procedure was considered converged when the relative change in total energy between consecutive iterations was below $10^{-5}$ and the norm of the global force vector of the system was less than or equal to $10^{-8}$~eV/\AA \cite{USTRZYCKA2024104118,Aligayev_2025}.
Next, the systems were equilibrated for 100~ps at 300~K using the isothermal-isobaric (NPT) ensemble, with a thermostat time constant of 100~fs.
The refractory WTaCrV alloy was constructed from the initial (001) W configuration by randomly substituting 75\% of the W atoms with Ta, Cr, and V atoms in equal amount.
The resulting WTaCrV configurations were subsequently subjected to the aforementioned energy minimization and thermal equilibration prior to the collision cascade simulations.

To simulate radiation-induced defect production in a simulation cell, a lattice atom was selected and assigned a high velocity, thereby generating a primary knock-on atom (PKA) and initiating a collision cascade.
The range of velocities investigated in this paper corresponds to the kinetic energies between 0.1 and 100~keV.
To obtain statistically representative set of cascades and minimize the influence of individual stochastic events, several independent cascade simulations were performed for each selected PKA energy (see Tab. \ref{table_MD}).
The PKA atoms were chosen from the shell of a sphere around the center of the simulation cell with a diameter of 0.6 cell dimensions and their velocity vectors were directed towards the center, in order to contain the cascade inside a cell and avoid self-interaction.
The PKA velocity directions were chosen in such a way as not to over-represent a single crystallographic direction.
During the ballistic and thermal spike stages, an adaptive time step was employed to improve computational efficiency while limiting the maximum atomic displacement to 0.02~\AA\ per integration step \cite{USTRZYCKA2024104118,SIMEONE2026122514,Aligayev_2025}.
For the equiatomic WTaCrV alloy, the projectile was randomly selected 
from the four constituent elements (W, Ta, Cr and V), with each species 
having an equal selection probability of 25\%, consistent with the 
nominal alloy composition. To account for the different atomic masses 
of the constituent elements, the initial velocity assigned to the 
selected PKA was calculated according to its atomic mass and the 
prescribed kinetic energy, \(E_{\mathrm{PKA}}=\frac{1}{2}m_i v_i^2\), 
where \(m_i\) and \(v_i\) are the mass and initial velocity of the 
selected atomic species $i$, respectively. Consequently, lighter elements such as Cr and V were assigned higher initial velocities than heavier 
elements such as W and Ta at the same PKA energy. This procedure 
ensured that all recoil events were initiated with the same prescribed 
kinetic energy while accounting for the species-dependent recoil 
velocities and their influence on collision kinematics and energy transfer during cascade evolution.

To reduce artificial reflections of pressure waves from the simulation boundaries and to remove excess heat generated by the cascade, a Nos\'e-Hoover thermostat was applied to a 5~\AA-thick boundary region of the periodically replicated simulation cell.
For high-energy cascades, electronic energy losses were taken into account for PKA energies of 10~keV and above.
This correction is necessary because a fraction of the kinetic energy of energetic atoms is transferred to the electronic subsystem rather than being available for elastic atomic collisions.
Neglecting this contribution can therefore lead to an overestimation of the energy retained by energetic recoils and consequently affect the predicted cascade dimensions and defect production.
Electronic stopping powers were obtained from the Stopping and Range of Ions in Matter (SRIM) calculations using the default stopping parameters for the constituent elements of the WTaCrV alloy \cite{Ziegler1985,ZIEGLER20101818}, and then used to account for the energy dissipated by sufficiently energetic atoms through interactions with the electronic subsystem during cascade evolution.

Analysis of the MD simulations, including polyhedral template matching, dislocation extraction algorithm, surface construction and Wigner-Seitz defect analysis, as well as structure visualization is carried out using OVITO \cite{ovito}.

The replaced atoms population was determined using the following OVITO workflow. First, the local structural environment of each atom in the relaxed post-cascade configuration was identified using the Polyhedral Template Matching (PTM) method to identify structure types, and the atoms that were identified as non-BCC were removed. For the remaining atoms, their positions were then compared with their corresponding positions in the optimized pre-cascade configuration using the Displacement Vectors modifier. Atoms for which the displacement magnitude was negligible, were filtered out, as their position changed marginally, on par with thermal oscillations. The remaining atoms, as such, represent the number of atoms that were replaced as a result of the collision cascade.

\begin{table}[t!]
    \caption{Summary of cascades features generated by the MD simulations. ($a_{bcc}= 0.316$ nm). all simulation were performed at $300K$.}
    \begin{tabular}{cccccc}
    \hline
       $E_K$ (keV)   &  Time (ps) & runs  & Size (in $a_{bcc}$) & atoms in the box  $N_B$\\
       \hline
       $0.1$ & $10$ & $25$ & $(52, 53, 54)$     & $3.0 \times 10^5$  &    \\
       $1$   & $10$ & $25$ & $(52, 53, 54)$     & $3.0 \times 10^5$  &     \\
       $3$   & $10$ & $25$ & $(52, 53, 54)$     & $3.0 \times 10^5$  &     \\
       $5$   & $10$ & $25$ & $(52, 53, 54)$     & $3.0 \times 10^5$  &     \\
       $10$  & $10$ & $25$ & $(52, 53, 54)$     & $3.0 \times 10^5$  &     \\
       $15$  & $10$ & $10$ & $(70, 71, 72)$     & $7.0 \times 10^5$  &     \\
       $20$  & $10$ & $10$ & $(70, 71, 72)$  & $7.0 \times 10^5$   &     \\
       $30$  & $10$ & $10$ & $(120, 121, 122)$  & $3.5 \times 10^6$   &     \\
       $50$  & $16$ & $10$ & $(120, 121, 122)$  & $3.5 \times 10^6$   &     \\
       $100$ & $35$ &  $5$ & $(170, 171, 172)$ & $1.0 \times 10^7$  &     \\
       \hline
    \end{tabular}    
    \label{table_MD}   
\end{table}

\section{Results}
\label{sec:results}

\subsection{Cascade evolution and morphology as a function of PKA energy}

\begin{figure}
    \centering
    \includegraphics[width=\linewidth]{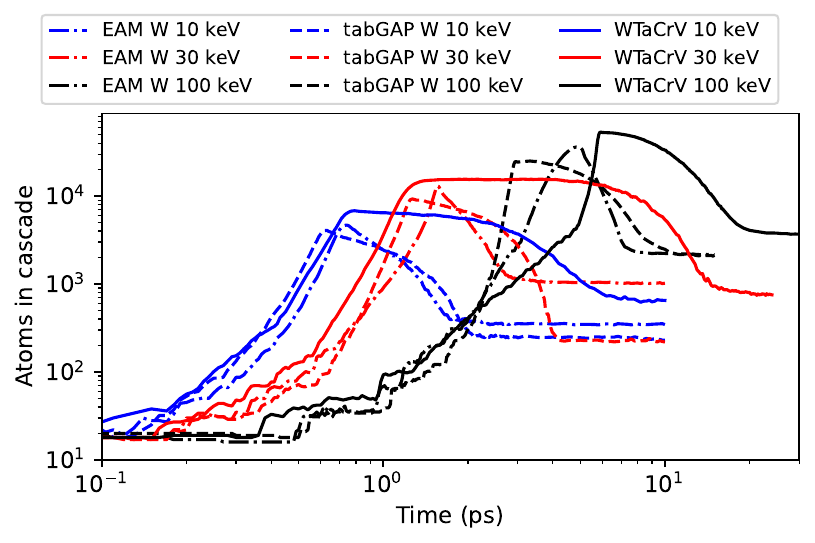}
    \caption{Comparison of cascade evolution for different PKA energies (denoted by color) in pure W and WTaCrV alloy. For W, results of two different potentials - EAM and tabGAP - are presented.}
    \label{fig:cascade-evolution}
\end{figure}

The number of atoms participating in the collision cascade as a function of simulation time is shown in Fig.~\ref{fig:cascade-evolution} for representative examples of PKA energies of 10, 30, and 100~keV.
Independent of the interatomic potential and material composition, the cascade evolution can be divided into three characteristic regimes: (i) the initial ballistic phase, (ii) the thermal-spike or heat-spike phase, and (iii) the subsequent recombination and relaxation phase.
The characteristic times associated with these regimes are not fixed but grow systematically together with PKA energy.
In particular, increasing the PKA energy extends the ballistic evolution and delays both the maximum cascade activity and the subsequent recovery of the damaged region \cite{Aligayev_2025}.

During the ballistic phase, the kinetic energy initially assigned to the PKA is rapidly transferred to neighboring lattice atoms through a sequence of energetic collisions.
Consequently, the number of atoms participating in the cascade increases rapidly from its initial value.
At 10~keV, this stage is relatively short and the cascade develops within the first fraction of a picosecond.
Increasing the PKA energy to 30 and further to 100~keV produces a longer and more spatially extensive collision sequence, as more kinetic energy is available for atomic displacements.
The differences among the potentials remain relatively modest during the early part of the cascade but become increasingly pronounced as the cascade develops.

The ballistic regime is followed by the heat-spike phase, characterized by a rapid increase in the number of participating atoms and the formation of a highly disordered, transiently excited region.
The magnitude and position of the maximum in time strongly depend on the PKA energy.
At 10~keV, the maximum cascade activity occurs at approximately $0.7$--$1$~ps and involves several thousand atoms.
At 30~keV, the maximum shifts toward approximately $1$--$2$~ps and reaches approximately $10^{4}$ atoms.
The 100~keV cascades exhibit substantially more complex behavior, with maxima occurring several picoseconds after cascade initiation and involving several tens of thousands of atoms.

Following the maximum cascade activity, the system enters the recombination and relaxation phase.
During this stage, the highly excited cascade region cools and a large fraction of the displaced atoms return to lattice sites (not necessarily the original ones, participating in the process of recombination), undergo vacancy-interstitial recombination, or, at higher temperatures, form higher-dimensional defects, such as stacking faults, at the same time contributing to the growing voids.
This process produces the pronounced decrease in the number of cascade-participating atoms after the heat-spike maximum.
The curves eventually approach a quasi-stationary value, indicating that the rapid cascade-induced recovery process has largely terminated.
The relaxation time increases with PKA energy: the 10~keV cascades recover within a few picoseconds, whereas the 30 and particularly the 100~keV cascades exhibit considerably longer relaxation times.
It is important to emphasize that the values in Fig.~\ref{fig:cascade-evolution} represent the number of atoms identified as participating in, or remaining displaced following, the cascade and should not be interpreted directly as the number of permanent point defects.

For pure W, simulations with the use of tabGAP potential are compared to those performed by using a well-know interatomic potential based on EAM method \cite{Marinica_2013}.
Both potentials reproduce the same overall sequence of ballistic excitation, heat-spike formation, and subsequent recovery, but quantitative differences are present in the magnitude and time of the heat-spike maximum, as well as heat-spike duration: EAM potential produces cascades that occur later, involve more atoms but last shorter than the cascade produced with tabGAP potential.
These differences become more pronounced as the PKA energy increases.

This observation highlights the sensitivity of cascade evolution to the description of the interatomic interactions, particularly under the strongly non-equilibrium conditions generated during the thermal spike.
The short-range interaction becomes especially important during the ballistic stage because energetic collisions can bring pairs of atoms to separations considerably smaller than those sampled under equilibrium conditions.
The ZBL screened Coulomb potential provides the strongly repulsive interaction required in this regime.
The tabGAP description without an explicit ZBL contribution relies on the short-range behavior contained in the machine-learning potential itself, whereas inclusion of the ZBL interaction imposes the physically motivated screened nuclear repulsion at sufficiently small interatomic separations.

For WTaCrV, the pronounced response obtained with the tabGAP+ZBL description at high PKA energy also suggests that the combination of short-range repulsive interactions and chemical complexity influences the evolution of energetic cascades.
In a chemically disordered multi-component alloy, differences in atomic masses and local atomic environments introduce additional variations in momentum transfer and collision sequences compared with elemental W.
Such effects can modify the spatial distribution of deposited energy and the subsequent recombination kinetics.

\begin{figure}
    \centering
    \includegraphics[width=\linewidth]{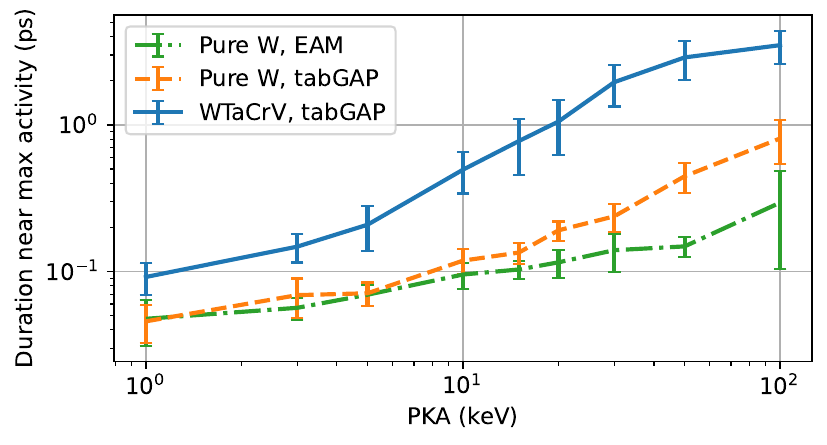}
    \caption{Comparison of heat-spike duration near maximum activity in picoseconds as a function of PKA in keV for pure W and WTaCrV equimolar alloy. Solid lines represent the mean values at each PKA, with error bars indicating standard deviation. For W, results of two different potentials - EAM and tabGAP - are presented.}
    \label{fig:peak-width}
\end{figure}

Fig. \ref{fig:peak-width} shows the duration of the heat-spike near it's maximum as a function of PKA energy for pure W described by the tabGAP and EAM potentials and for WTaCrV HEA described by the tabGAP potential.
Here, the full width at 80\% height above relaxed post-cascade state is used as a measure of the heat-spike duration near maximum, i.e., the characteristic time interval over which the cascade remains in its maximally excited and highly disordered state.
The lines represent the mean values of different realizations at each PKA energy with the error bars showing standard deviation.
At low PKA energies, the thermal-spike duration near maximum exhibits only a weak dependence on PKA energy.
For pure W, both the EAM and tabGAP descriptions yield relatively short durations near maximum, of the order of a few hundredths of a picosecond at the lowest energies.
The similarity between the two potentials in this regime indicates that low-energy cascades are relatively localized and short-lived.
The WTaCrV alloy exhibits systematically larger average durations near maximum than pure W, indicating a longer-lived highly disordered state.
This difference suggests that chemical disorder, together with the associated variation in atomic masses and local atomic environments, modifies the redistribution of the PKA energy and the subsequent relaxation kinetics.

A pronounced change in behavior becomes apparent at PKA energies above approximately 10~keV.
Beyond this energy, the average thermal-spike duration near maximum increases more rapidly with increasing PKA energy, particularly for WTaCrV.
For pure W, the tabGAP results also show a progressively increasing duration and deviate increasingly from the EAM predictions at high energies.
At 100~keV, the average duration obtained with tabGAP approaches the picosecond range, whereas the EAM calculation remains considerably shorter.
This increasing divergence indicates that the predicted lifetime of the highly excited cascade state becomes increasingly sensitive to the interatomic description as the deposited energy and cascade dimensions increase.
WTaCrV shows stronger PKA energy dependence of duration near maximum between 5~keV and 30~keV compared to pure W, showing a substantially steeper increase, with the characteristic duration reaching the picosecond range at higher PKA energies.
On average, a noticeable increase in the dispersion among individual cascade realizations is also evident in WTaCrV compared to pure W.
This larger variability indicates that high-energy cascade evolution becomes increasingly sensitive to the particular PKA direction and local chemical environment sampled by each trajectory.

\begin{figure*}
    \centering
    \begin{overpic}[width=0.4\textwidth]{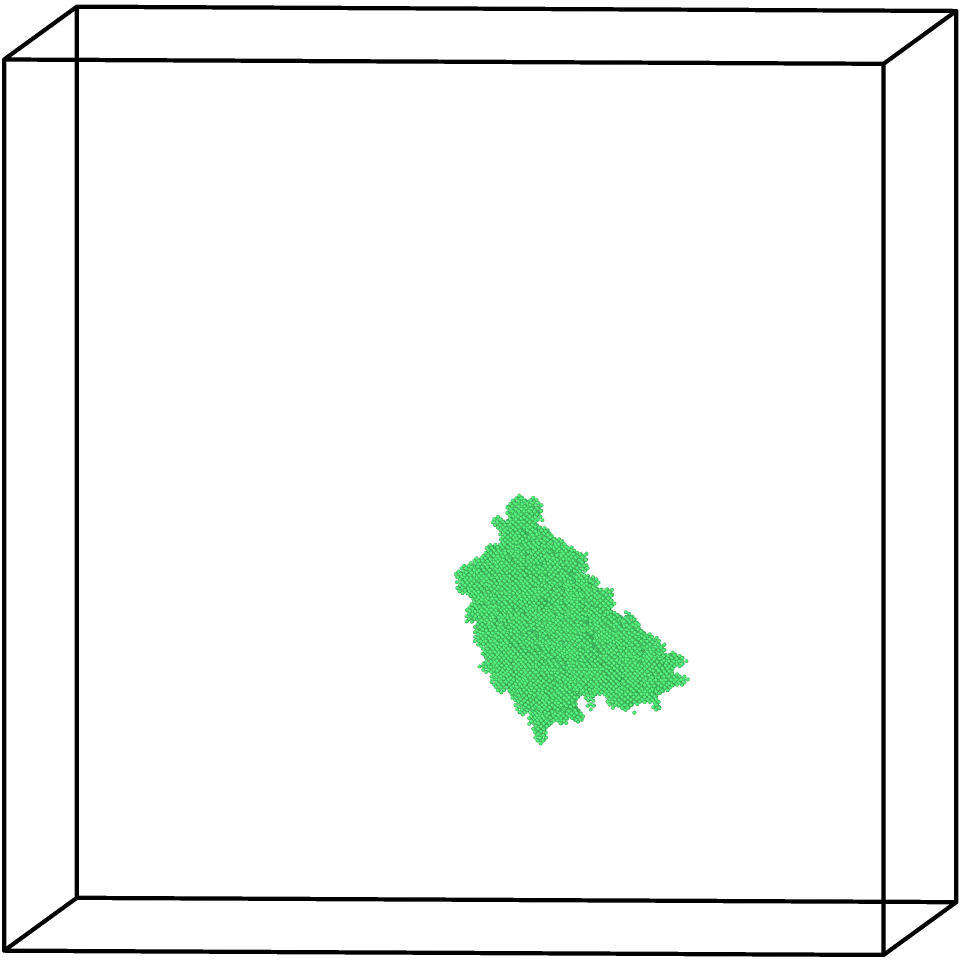}
        \put(10,80){\large a)}
    \end{overpic}  \quad \quad
    \begin{overpic}[width=0.4\textwidth]{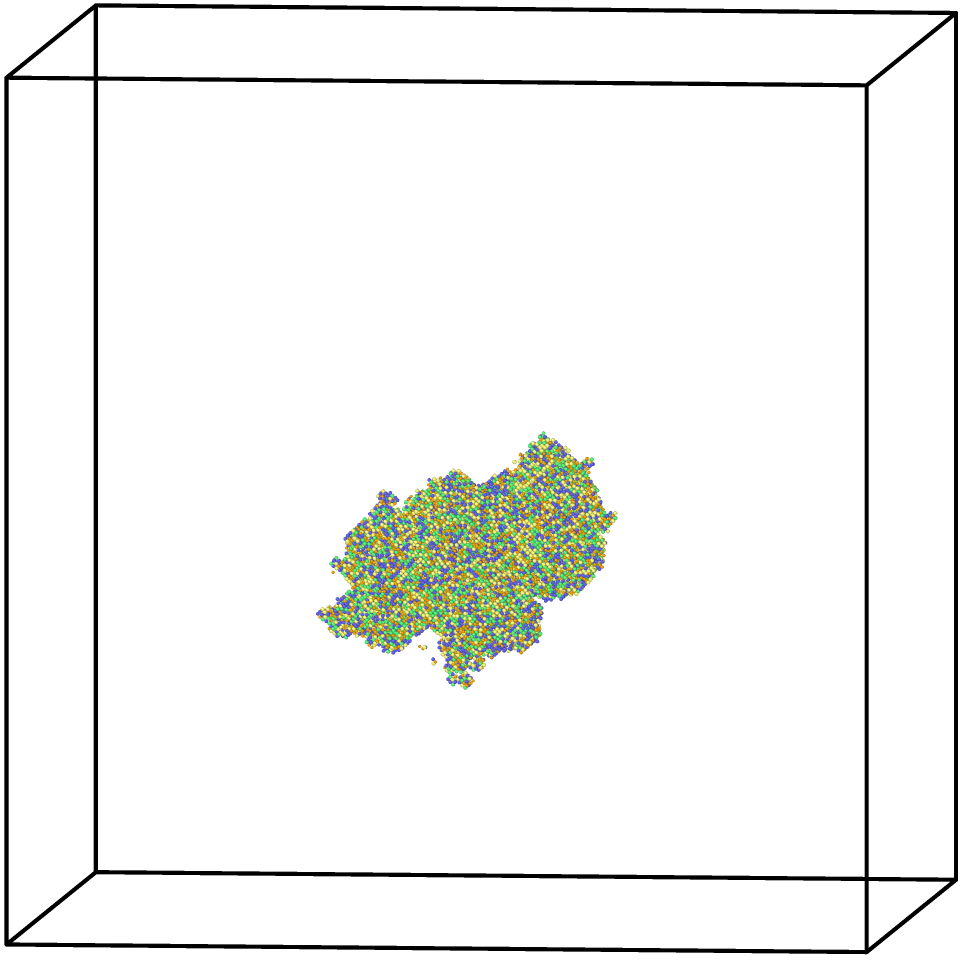}
        \put(15,80){\large b)}
    \end{overpic} 
    \begin{overpic}[width=0.4\textwidth]{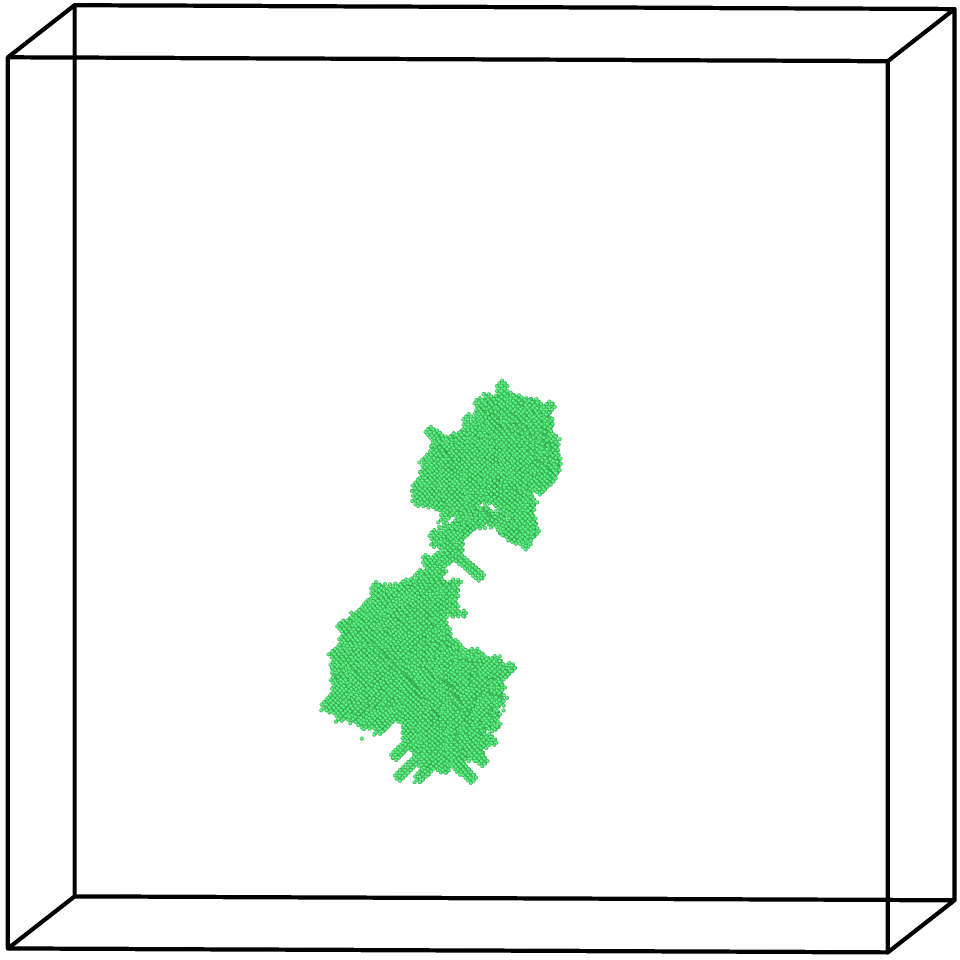}
        \put(15,80){\large c)}
    \end{overpic}  \quad \quad
    \begin{overpic}[width=0.4\textwidth]{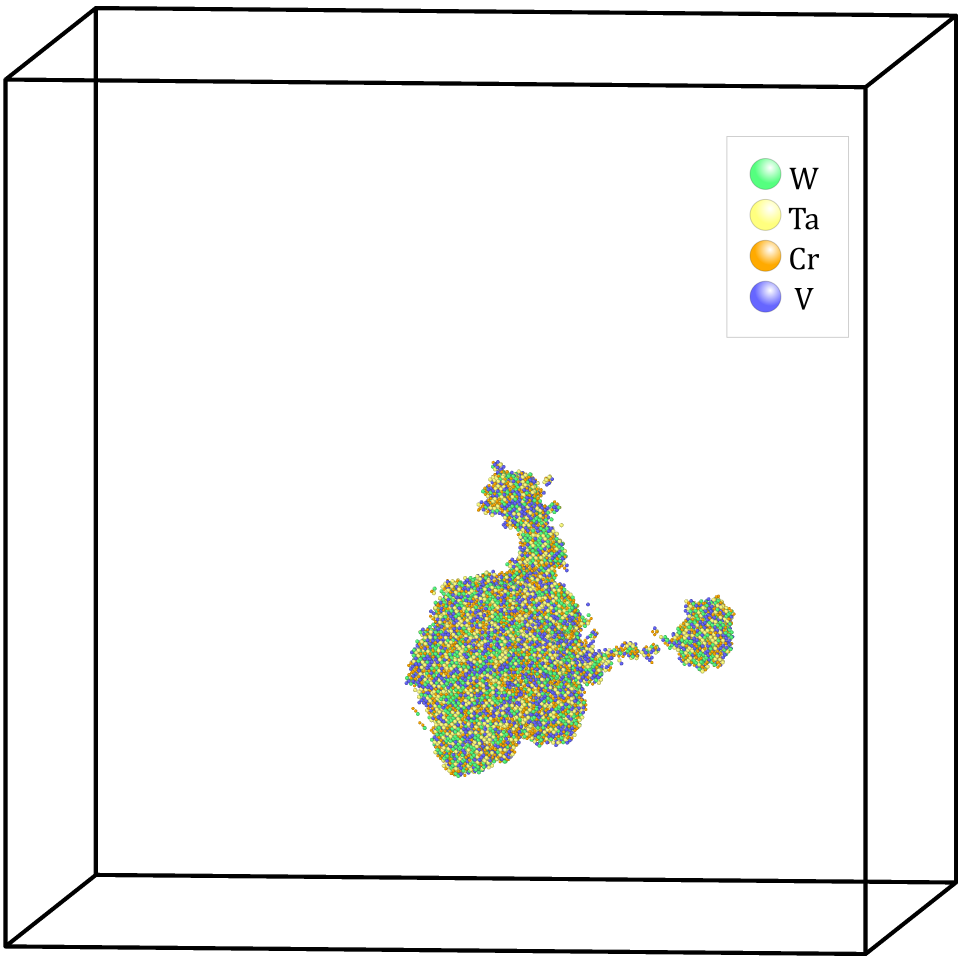}
        \put(15,80){\large d)}
    \end{overpic}
     \begin{overpic}[width=0.4\textwidth]{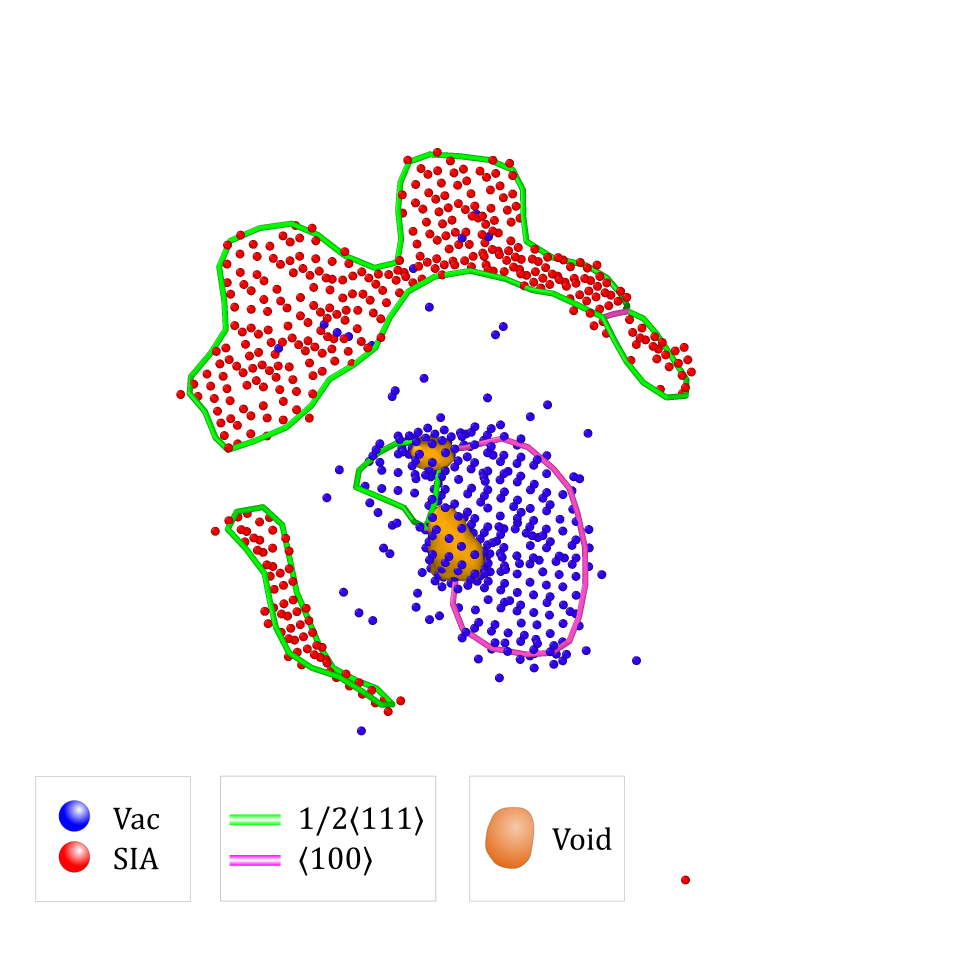}
        \put(15,80){\large e)}
    \end{overpic}  \quad \quad
    \begin{overpic}[width=0.4\textwidth]{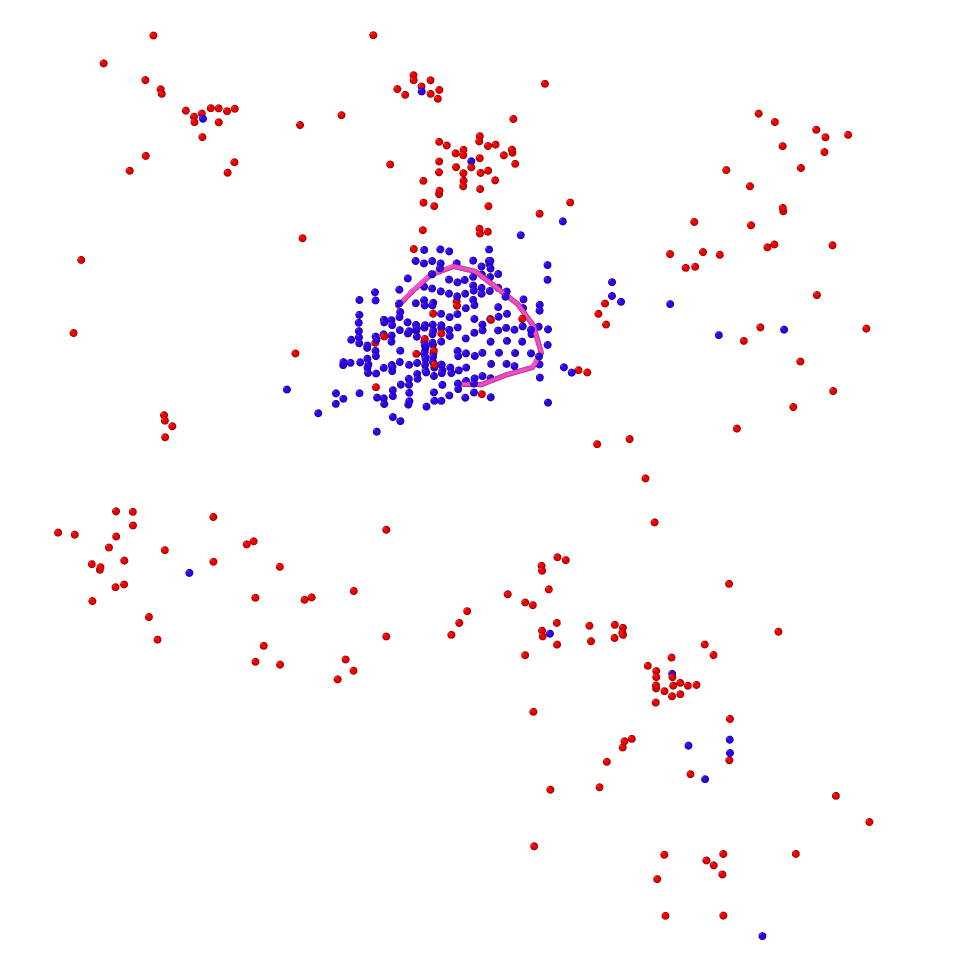}
        \put(15,10){\large f)}
    \end{overpic}
    \caption{Representative cascades at their maximums and defect morphology after the cascade for 100~keV PKA for pure W and WTaCrV obtained with tabGAP potential. Top row shows examples of non-split cascade in (a) pure W and (b) WTaCrV. Middle row shows examples of split cascade in (c) pure W and (d) WTaCrV. Bottom row shows the defect structure after the cascade in (e) pure W and (f) WTaCrV. The scale in (e) and (f) is consistent with each other, but different from the scale in (a-d).}
    \label{fig:visualization}
\end{figure*}

An important qualitative measure of the collision cascade is the subcascade formation.
A situation where small quantities of kinetic energy are transferred to neighboring atoms, which transfer it to their neighbors and so on, thus creating a uniform volume around the PKA trajectory is not the only one possible.
Two alternative situations can occur: (i) PKA transfers a significant amount of kinetic energy to one particle, which in turn starts it's own collision cascade, and (ii) at some point PKA atom with enough kinetic energy left moves through an "easy" crystallographic direction barely affecting neighboring atoms until it hits an atom in a region not affected by the original cascade.
In both of the situations, multiple volumes of the material are affected by one PKA, which are called subcascades \cite{LUNEVILLE201155,SIMEONE2026122514}.
Examples of singular cascade and split cascade are shown in Fig.~\ref{fig:visualization}(a-d).

\begin{figure}
    \centering
    \includegraphics[width=\linewidth]{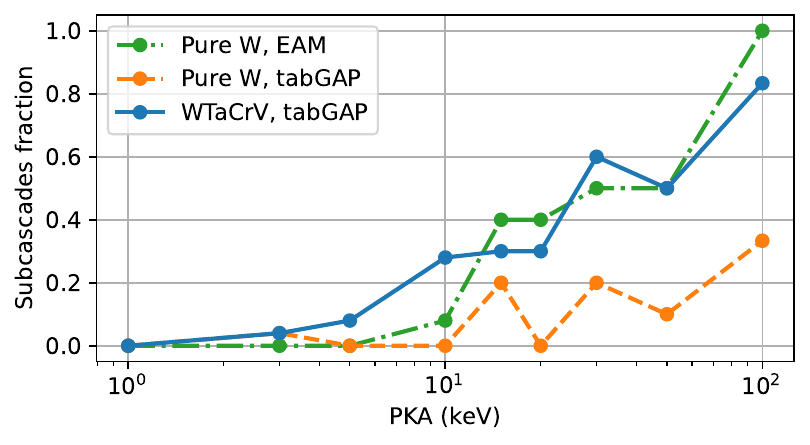}
    \caption{Fraction of the simulations in which cascade splits into multiple subcascades for pure W (simulated with EAM and tabGAP potentials) and WTaCrV as a function of PKA energy.}
    \label{fig:subcascades}
\end{figure}

Fig.~\ref{fig:subcascades} shows the fraction of simulations in which subcascades were observed at each PKA energy.
For pure W, EAM potential is prone to producing subcascades much more often than the tabGAP potential and at lower PKA energies.
Analytical estimations of the threshold energies for cascade splitting from Ref.~\cite{LUNEVILLE201155} suggest that the subcascades formation for W should occur above $\sim$50~keV.
Evidently, results produced with tabGAP potential are much more accurate compared to EAM potential. As such the decision have been made to leave out the analysis of results obtained with EAM potential in the rest of the paper for the sake of clarity.
When comparing results for W and WTaCrV with the same tabGAP potential, it is noticeable that in the HEA, subcascades form more often at 5~keV PKA energy and above.
Although subcascade formation is a stochastic process, chemical inhomogeneity, meaning atoms with different masses and sizes being present in the system, leads to a larger variety of atomic collisions, promoting the higher frequency of subcascade observations.

\begin{figure}[t]
    \centering
    \includegraphics[width=0.48\textwidth]{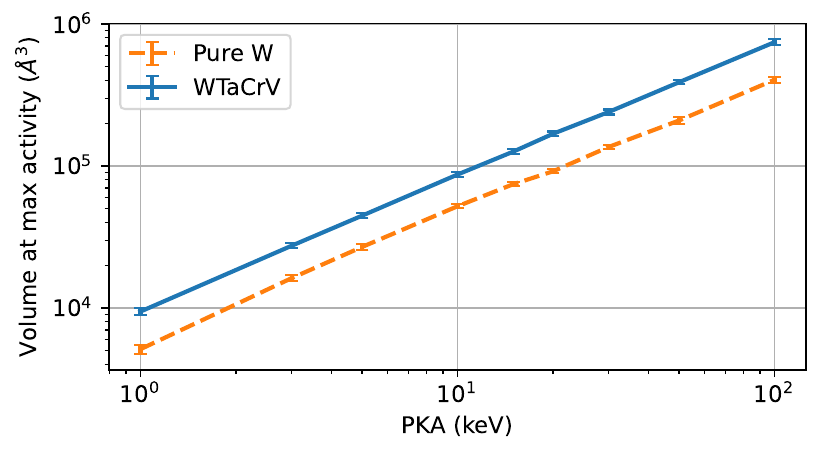}
    \caption{Comparison of volume in \AA$^3$ occupied by the cascade at its peak as a function of PKA in keV for pure W and WTaCrV equimolar alloy. Solid lines represent the mean values at each PKA with error bars indicating standard deviation.}
    \label{fig:peak-volume}
\end{figure}

Fig. \ref{fig:peak-volume} shows the maximum heat-spike volume as a function of PKA energy for pure W described by the tabGAP and EAM potentials and for the WTaCrV RHEA described by the tabGAP potential.
The heat-spike volume was evaluated at the time corresponding to the maximum cascade activity.
At this instant, atoms retaining the ideal BCC local environment were excluded, and the remaining non-BCC atoms were taken to define the transient structurally disordered region generated by the cascade.
A three-dimensional surface mesh was then constructed around these atoms using OVITO's\cite{ovito} Construct Surface Mesh modifier with alpha-shape algorithm\cite{stukowski2014jom}.

The heat-spike volume increases systematically with PKA energy both for pure W and WTaCrV, reflecting the progressively larger region affected as more kinetic energy is deposited into the lattice.
In case of WTaCrV, the volume of the cascade is systematically larger compared to pure W. This, again, is the effect of compositional complexity, more specifically, the presence of lighter elements in the mix (Cr and V), which require less kinetic energy to be removed from equilibrium lattice positions.

\begin{figure}[t]
    \centering
    \includegraphics[width=\linewidth]{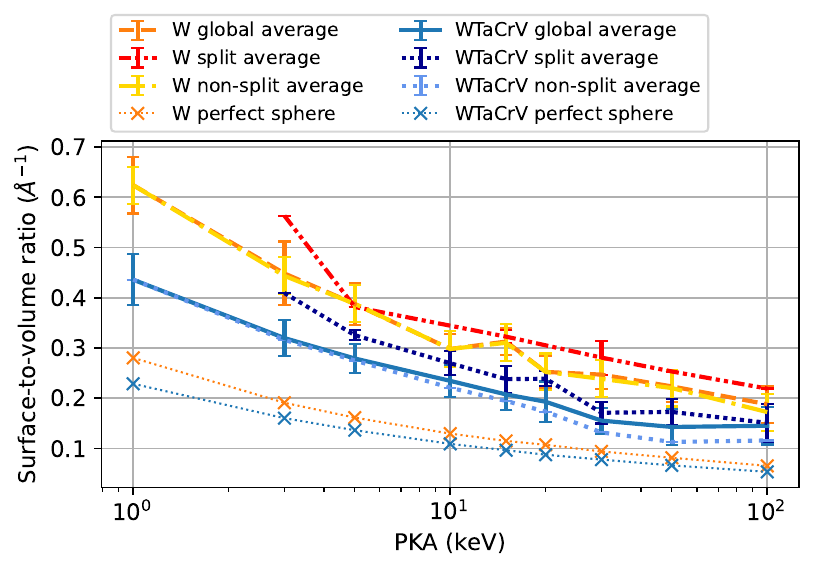}
    \caption{Ratios of surface to volume of the heat-spike during maximum activity as a function of PKA energy for pure W and WTaCrV. Apart from global average, averages for split cascades and non-split cascades for each alloy are presented. Surface-to-volume ratio of perfect spheres for the average volumes at corresponding PKA energies for W and WTaCrV are given for reference.}
    \label{fig:ratios}
\end{figure}

No significant deviations are seen for the volume at maximum heat-spike activity.
However, cascades may have different shape, regardless of splitting (as seen in Fig.~\ref{fig:visualization}).
As a measure of shape difference, Fig.~\ref{fig:ratios} presents the ratio of surface to volume as a function of PKA energy.
Due to sphere having lowest surface-to-volume ratio, the reference for perfectly spherical cascades corresponding to the average volumes at each PKA energy for W and WTaCrV are also presented in the figure.
All cascades have surface-to-volume ratio higher than that of a perfect sphere, meaning they all have non-spherical shapes.
When comparing ratios for the split and non-split cascades, the non-split cascades naturally have shapes closer to spherical, indicated by lower surface-to-volume ratios, than split cascades, which have higher surface-to-volume ratios for both pure W and WTaCrV.
This can mean that the split cascade, even with the identical geometric volume as a non-split cascade, can affect the larger region in the sample.
Taking it into account and combining it with the facts that the average volume of heat-spike at its maximum and the fraction of split cascades in WTaCrV are higher than in pure W, the damage in the alloy is spread over a larger volume of the sample.
This is visible in Fig.~\ref{fig:visualization}(e,f) when comparing defect visualization in pure W and WTaCrV presented at the same scale.

\begin{figure}
    \centering
    \includegraphics[width=\linewidth]{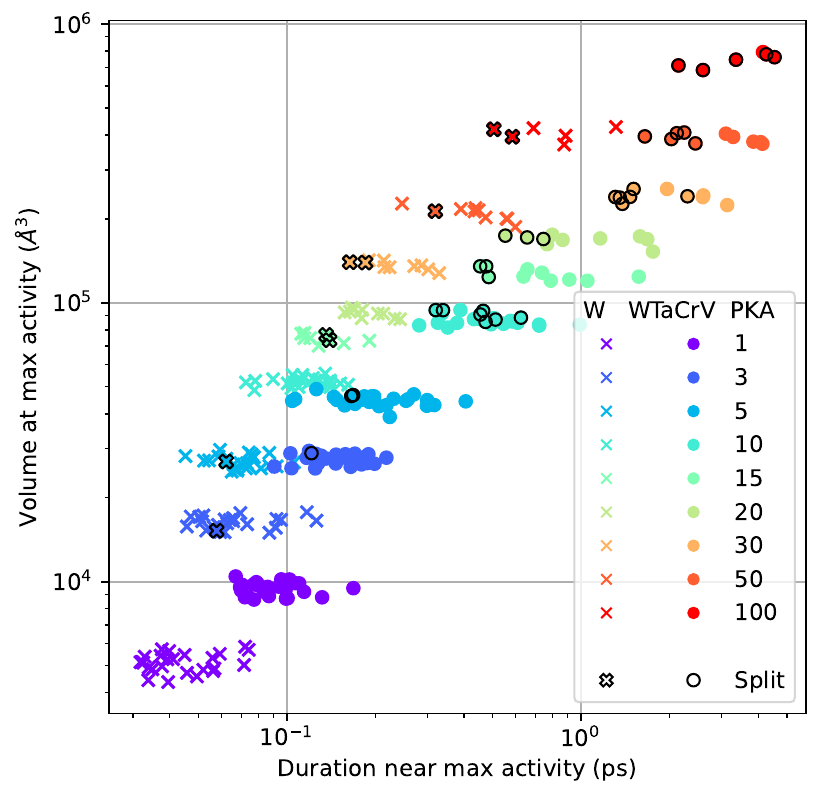}
    \caption{Volume occupied by cascade at its maximum activity as a function of duration of cascade near maximum activity for W and WTaCrV alloy simulated with tabGAP potential. Additionally, PKA energies are indicated by color, and markers for split cascades are highlighted by black borders.}
    \label{fig:correlation}
\end{figure}

Fig.~\ref{fig:correlation} shows correlations between heat-spike volume, duration, PKA energy, and cascade splitting.
Two more trends are visible in this graph, in addition to those identified in Figs.~\ref{fig:peak-width}, \ref{fig:subcascades} and \ref{fig:peak-volume}.
First, the duration of heat spikes is less affected than the volume of cascade at the peak with increasing PKA energy for pure W compared to the WTaCrV alloy.
Second, the cascades that are split into subcascades tend to last shorter for every PKA energy compared to the non-split cascades.
A possible explanation for this observation is that since split cascades have a higher surface area compared to non-split cascades, the heat exchange rate between the locally molten region affected by cascade and the non-affected region of the sample with lower temperature is higher, reducing the duration of the heat spike.

\subsection{Analysis of replaced atoms}

The analysis of replaced atoms captures the extent to which the collision cascade reorganizes the crystalline lattice, even when atoms ultimately recover a BCC local environment.
A displaced atom is defined as an atom that is still identified as belonging to a BCC environment after the cascade but whose final position differs from its original lattice position.
Such atoms therefore do not necessarily correspond to surviving vacancies or self-interstitial atoms.
Instead, they reflect atomic replacement sequences, local lattice rearrangements, and the net transport of atoms produced during the ballistic and thermal-spike stages.
This quantity is important because collision cascades can induce extensive atomic mixing and replacement without producing an equivalent number of point defects\cite{TERENTYEV2006119}.

\begin{figure}[t]
    \centering
    \begin{overpic}[width=0.46\textwidth]{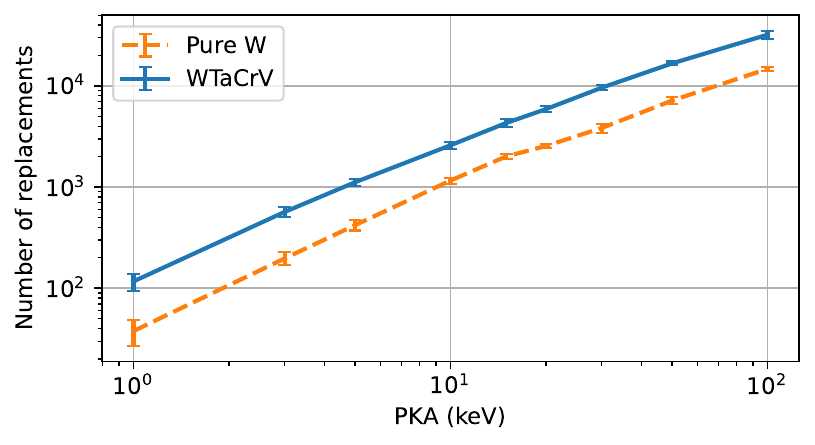}
        \put(90,15){\large a)}
    \end{overpic}
    \begin{overpic}[width=0.48\textwidth]{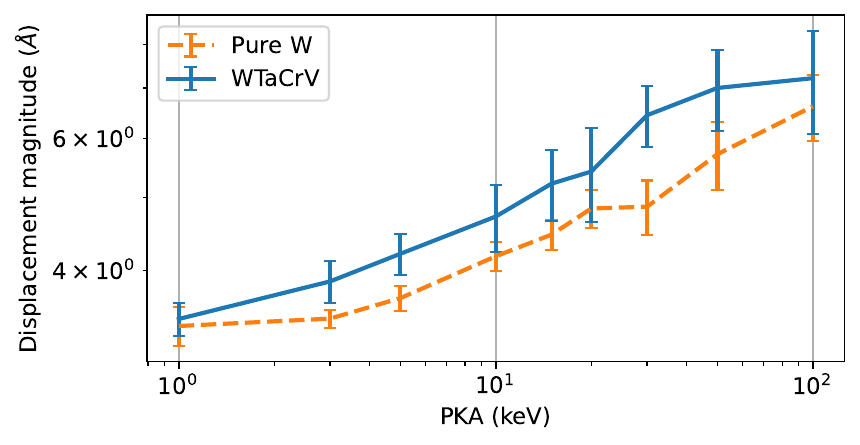}
        \put(90,15){\large b)}
    \end{overpic}
    \caption{Comparison of (a) number of atomic replacements and (b) average displacement magnitude  as a function of PKA in keV for pure W and WTaCrV equimolar alloy. Solid lines represent the mean values at each PKA with error bars representing standard deviation.}
    \label{fig:replacements}
\end{figure}

Fig. \ref{fig:replacements}(a) shows the average number of replaced BCC atoms as a function of PKA energy for pure W and the WTaCrV HEA.
For both materials, the number of replaced atoms increases strongly and monotonically with increasing PKA energy, reflecting the progressively larger total number of atoms involved in ballistic collision sequences and thermal-spike-induced atomic rearrangements.
At the lowest investigated energies, pure W exhibits only several tens of replaced atoms, whereas the WTaCrV alloy already shows a substantially larger replaced population.
With increasing PKA energy, this difference persists over the entire energy range.
At 10~keV, the average number of replaced atoms reaches approximately $10^{3}$ in pure W and several $10^{3}$ in WTaCrV, while at 100~keV both systems contain replaced atoms on the order of $10^{4}$, with the alloy remaining systematically above pure W.

The systematically larger replaced atoms population in WTaCrV indicates that collision cascades produce more extensive atomic rearrangement in the chemically complex alloy than in elemental W.
This observation is consistent with the longer thermal-spike durations and larger heat-spike volumes obtained for WTaCrV, as seen in Figs.~\ref{fig:peak-width} and \ref{fig:peak-volume}.
The combined results suggest that the deposited PKA energy is distributed over a larger number of atoms in the alloy affecting a larger region of the material.
The strong atomic-mass contrast among W, Ta, Cr, and V, together with variations in local chemical environments, produces heterogeneous momentum-transfer pathways during the ballistic stage.
These effects can enhance replacement collision sequences and ballistic mixing, allowing atoms to undergo significant number of replacements even when the local BCC structure is subsequently recovered during cascade cooling.

Fig. \ref{fig:replacements}(b) presents the average displacement magnitude of the BCC atoms displaced from their original lattice positions after the collision cascade, but returning to another BCC lattice position, as a function of PKA energy.
Starting at the lowest PKA energy, 1~keV, average displacement distance is higher in WTaCrV compared to pure W.
The gap between the two increases immediately at 3~keV, and remains more or less the same until 30~keV, where it reaches its maximum, and the gap starts decreasing again.
Such change in the difference is caused by the plateauing of the curve for WTaCrV, while for pure W it is gradually increasing.
This trend is similar to the one observed for the duration of heat spike near its maximum, shown in Fig.~\ref{fig:peak-width}.

\subsection{Analysis of defects}

\begin{figure*}
\centering
\begin{overpic}[width=0.46\textwidth]{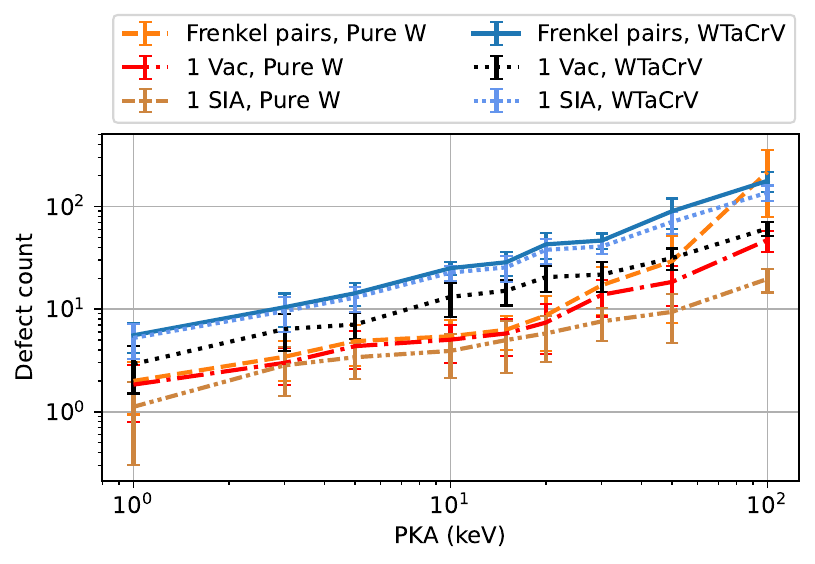}
        \put(90,15){\large a)}
    \end{overpic}
    \begin{overpic}[width=0.46\textwidth]{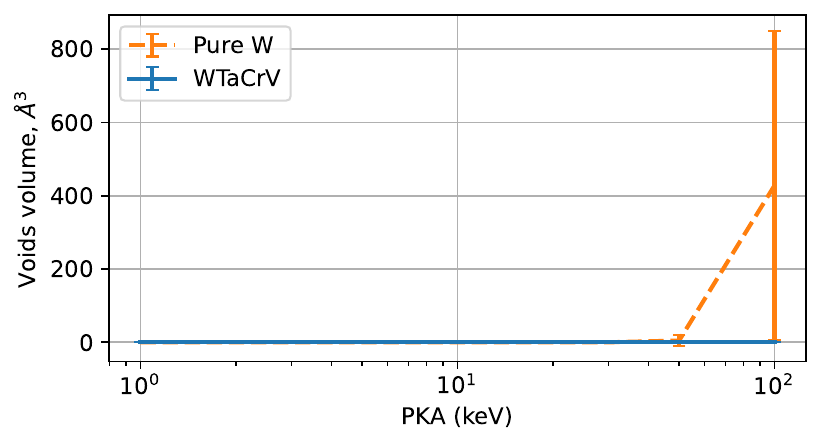}
        \put(20,15){\large b)}
    \end{overpic}
    \begin{overpic}[width=0.46\textwidth]{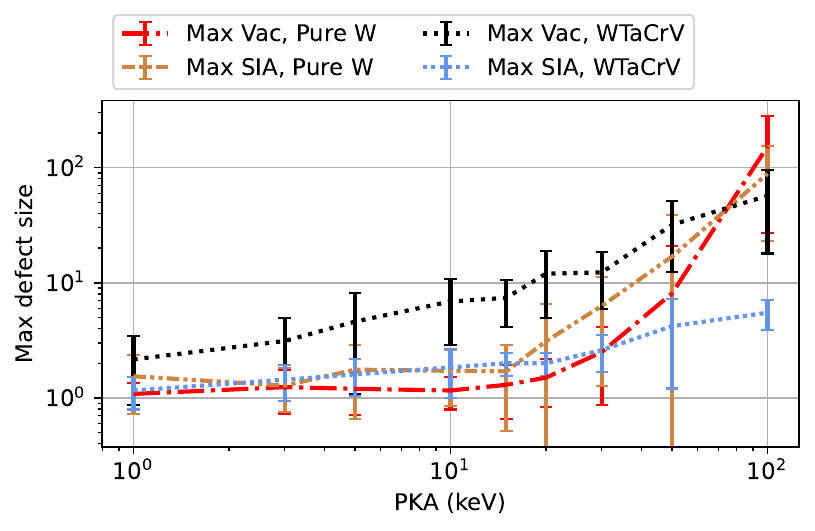}
        \put(90,15){\large c)}
    \end{overpic}
    \begin{overpic}[width=0.46\textwidth]{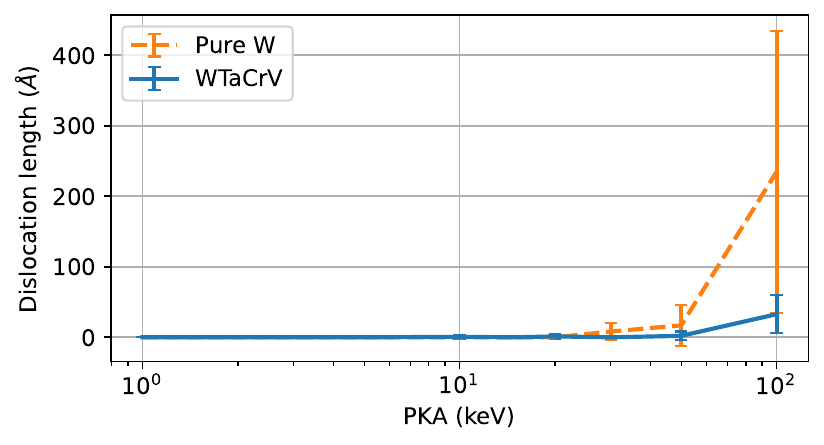}
        \put(20,15){\large d)}
    \end{overpic}
\caption{Defect analysis in pure W and WTaCrV as a function of PKA energy: (a) Frenkel pairs and number of single vacancies as well as SIAs, (b) number of defects in largest vacancy and sia clusters, (c) void volume, (d) total dislocation length. Solid lines represent the mean values at each PKA with error bars representing standard deviation.}
\label{fig:defects}
\end{figure*}

Complex analysis of 1D, 2D and 3D defects is presented in Fig.~\ref{fig:defects}.
Frenkel pairs, presented in Fig.~\ref{fig:defects}(a) have been calculated by a Wigner-Seitz defect analysis in OVITO.
Due to the nature of this analysis, which counts the occupancy of the volume belonging to each atomic site in the undisturbed reference structure (pre-cascade), the yielded number of Frenkel pairs does not strictly represent a pair of isolated defects, but also counts vacancies that constitute three-dimensional voids and self-interstitial atoms that form a stacking fault, for example.
Therefore, the number of such singular defects in each material is presented in Fig.~\ref{fig:defects}(a) separately.

For both materials, the total trend of growing number of defects with increasing PKA energy is observed.
The number of isolated defects (both vacancies and SIAs) is also lower than the total number of Frenkel pairs in both materials, indicating that defects may form defect clusters.
The relation between the monovacancies and single SIAs, however, is different in the two materials.
For pure W, number of singular SIAs increases the slowest with PKA energy, starting to lag progressively behind the total number of Frenkel pairs above 20~keV.
This suggests that SIAs can form defect clusters and even defect cluster planes, which is confirmed by a dislocation analysis shown in Fig.~\ref{fig:defects}(d), where dislocations are observed in pure W above 20~keV.
In WTaCrV, noticeable amount of dislocation is observed only at 100~keV, and their total length is on average 15 times smaller than in pure W.
This is consistent with Fig.~\ref{fig:defects}(a), which shows that almost all SIAs in WTaCrV are isolated SIAs, and with Fig.~\ref{fig:defects}(b), which shows that at 100~keV the average number of SIAs in the biggest cluster in WTaCrV (6 sites) is much smaller compared to that in pure W (89 sites).

The number of monovacancies is systematically lower than the number of isolated SIAs in WTaCrV, with minor increase in the discrepancy with increasing PKA energy.
For pure W, the number of monovacancies is systematically higher compared to isolated SIAs.
It shows monotonous increase, but the relative fraction in relation to total Frenkel pairs drops at 50~keV and further at 100~keV -- a lag occurring later compared to SIAs.
In Fig.~\ref{fig:defects}(c) it can be seen that for pure W, the rest of the vacancies accounted for by the Frenkel pair analysis are present in the form of voids.
For WTaCrV, no voids are observed at all.
The size of the maximum cluster of voids in WTaCrV (see Fig.~\ref{fig:defects}(b)) is $\sim$2.6 times lower compared to pure W (57 and 153 atomic sites, respectively) but is still significantly larger than the medium-sized di- or tri-vacancy cluster.
The nature of the biggest clusters in WTaCrV, however, is not volumetric voids, but the $\langle100\rangle$ dislocations instead.

Cascade splitting effect on dislocation length and void volumes at 100~keV is as follows. For pure W, voids can form in both cases, but the volume of the void introduced by singular cascades, as compared to a split cascade, is $\sim$2.5 times higher.
The length of $\nicefrac{1}{2}\langle111\rangle$ dislocations is higher for non-split cascades compared to split cascades in both pure W and WTaCrV times $\sim$1.8 and $\sim$3.5, respectively.
The effect on the formation of $\langle100\rangle$ dislocations is opposite in two materials: in pure W, $\langle100\rangle$ dislocations were observed only for non-split cascades, whereas in WTaCrV they were observed only for split cascades.
Due to the small number of simulations -- 6 for each material --, the numbers presented in this paragraph should be taken with a grain of salt.
The general trend, for voids and $\nicefrac{1}{2}\langle111\rangle$ dislocations dislocations, however, suggests that 2D and 3D defects are formed more easily when the cascade is more localized, latter being more frequently observed in W than in WTaCrV.

\begin{figure*}
    \centering
    \begin{overpic}[width=0.49\textwidth]{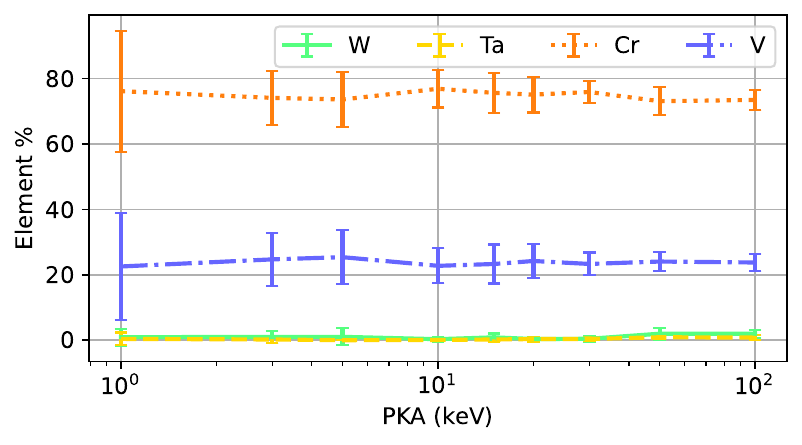}
        \put(1,50){\large a)}
        \end{overpic} 
    \begin{overpic}[width=0.49\textwidth]{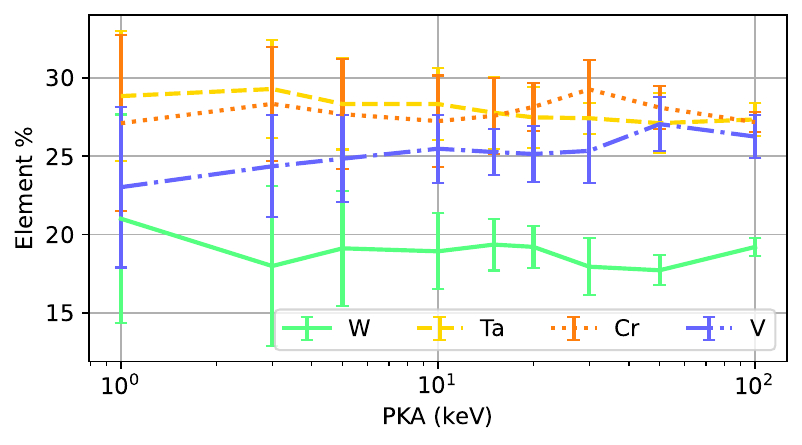}
        \put(1,50){\large b)}
        \end{overpic} 
    \caption{Elemental distribution of (a) self-interstitials and (b) vacancy/void neighbors after a collision cascade as a function of PKA energy in pure W and WTaCrV.}
    \label{fig:elemental-distribution}
\end{figure*}

Fig.~\ref{fig:elemental-distribution} shows elemental distribution in defect-related parts of the simulation cells for WTaCrV alloy.
The chemical composition of SIAs was calculated by taking the composition of atoms in the sites with increased occupancy.
It is shown in Fig.~\ref{fig:elemental-distribution}(a).
It can be seen that the majority of SIA population consists of Cr -- the smallest and the most lightweight atom in the alloy.
Second most present element is V -- second most lightweight and second smallest atom.
Both heavy elements, Ta and W, practically do not form SIAs.
These results are in accordance with the results for non-equiatomic W$_{38}$Ta$_{36}$Cr$_{15}$V$_{11}$\cite{chen2023journalofnuclearmaterials}.
This is also consistent with the work in Ref.~\cite{wei2024actamaterialia}, which identifies the critical role of V, the most lightweight element in various W-based HEAs investigated in that work, as being the one mitigating the growth of voids.

For vacancies, the calculation is less straightforward than for SIAs and was done in the following way: disturbed atoms, i.e. ones not belonging to ordered BCC structure were selected from the cell; next, among them atoms in the neighborhood of SIAs were selected and filtered out; the remaining pool of atoms was considered related to vacancies, voids or void planes.
Tungsten is the element that is most rarely found near the vacancy-related defects, while Ta and Cr are found the most often.
V shows a trend in being more often observed around vacancy-related defects with increasing PKA energy, reaching the values comparable to that of Ta and Cr at 50~keV.

As a note, presenting dislocation length instead of dislocation density in Fig.~\ref{fig:defects}(d) is a deliberate choice, motivated by the fact that the size of the cells for the simulations of different PKAs varies significantly, thus being an extra variable in the equation for the calculation of the dislocation density.
For the 100~keV PKA, the dislocation density is equal to approximately $2\times10^{14}$~m$^{-2}$.

Defect-wise, homogeneous W and chemically complex WTaCrV -- having two light and two heavy elements -- show two distinct microstructures as a result of high-energy collision cascades.
In pure W, collision cascades introduce voids or large $\langle100\rangle$ dislocations near the position of cascade origin, redistributing the displaced atoms mainly in a form of extra planes, surrounded by $\nicefrac{1}{2}\langle111\rangle$ dislocations, with the rest forming individual SIAs.
The effect is more noticeable in case of more localized heat spikes, which are the majority of observed cascades in pure W, and constitute the minority the alloy.
In WTaCrV, the cascade may cause the formation of small $\langle100\rangle$ dislocation, but mainly the damage is distributed throughout the whole affected volume in form of monovacancies or occasional medium-sized vacancy clusters and numerous isolated SIAs.

In summary, the results presented in this paper indicate that the systematically larger cascade volumes, longer characteristic thermal-spike durations, and increased atomic replacement activity observed 
in WTaCrV suggest that the initially deposited kinetic energy is 
redistributed through a larger number of atomic interactions and over
a more spatially extended region. Such behavior is consistent with the 
combined influence of atomic-mass contrast, local chemical disorder, 
and variations in the interatomic interactions among the constituent 
elements, which can modify cascade trajectories and energy-transfer 
pathways during the ballistic and relaxation stages. In particular, 
the higher probability of subcascade formation in WTaCrV provides a 
plausible explanation for the reduced development of concentrated 
defect structures, since the fragmentation of a primary cascade into 
spatially separated regions may limit the local accumulation of 
defects necessary for the nucleation and growth of extended dislocation 
structures and vacancy clusters. This interpretation is supported by 
the observed tendency of more localized, non-split cascades to produce 
larger extended defects, particularly in pure W.
The 
preferential occurrence of Cr and V atoms in self-interstitial 
configurations further suggests that the lighter constituent elements in the WTaCrV alloy 
play an important role in accommodating cascade-induced lattice 
disorder, potentially favoring the retention of dispersed point defects 
over their organization into extended structures. 

Importantly, the 
larger number of atomic replacements and the persistence of isolated 
point defects in WTaCrV demonstrate that the reduced formation of 
dislocations and voids does not necessarily correspond to a reduction 
in the total amount of primary damage, but rather to a modification of 
its spatial distribution and structural organization.
While not considering the long-term irradiation effects, such as defect 
migration, clustering, and accumulation under sustained irradiation, the presented results show that compositional complexity in WTaCrV 
modifies cascade energy redistribution and suppresses the formation of certain extended defect structures even during the primary irradiation damage. Apart from an atomistic basis for 
understanding how multi-component alloying may influence primary 
radiation-damage morphology, our findings offer useful information for future 
multiscale investigations of irradiation-induced microstructural evolution in refractory alloys\cite{SIMEONE2026122514}.

\section{Conclusions}
\label{sec:conclusions}

In this work, molecular dynamics simulations were used to investigate the energy-dependent evolution of collision cascades in pure W and the WTaCrV refractory high-entropy alloy over a broad range of PKA energies.
The analysis focused on the morphology of the simulation cell region affected by cascade and the defect makeup introduced by the cascade.

At low energies, EAM and tabGAP provide comparable descriptions of cascades in pure W, whereas their predictions increasingly diverge above 10~keV. In particular, tabGAP predicts a stronger increase in the thermal-spike lifetime and lower probability of subcascade formation, more compatible with analytical model~\cite{LUNEVILLE201155}, highlighting the importance of an accurate description of interatomic interactions under highly energetic collision conditions.
PKA energy increase leads to the formation of disordered regions that are larger in volume and longer-lived, with a growing probability of cascades splitting into multiple subcascades. Compared with pure W, the WTaCrV alloy exhibits larger heat spike volumes, longer thermal-spike durations, and a greater number of atomic replacements, with the average displacement of the atoms and the heat spike duration plateauing at high PKA energies.
In contrast to larger heat spike volume, the damage in WTaCrV is spread over a larger region of the simulation cell due to higher probability of the subcascade formation, signifying the crucial role of the latter for the mesoscale irradiation modeling. 

Another difference between pure W and WTaCrV concerns the formation of defects.
Point defects are much more prevalent in WTaCrV as compared to the pure W, where extended defects, such as voids and dislocations are formed instead. At PKA energies of approximately 50 keV and above the total length of dislocations in pure W is five times higher compared to WTaCrV.
No volumetric voids are observed in WTaCrV, while a sharp increase in their presence in pure W is observed for 100~keV PKA.
Again, the subcascade splitting plays a huge role, as $\nicefrac{1}{2}\langle111\rangle$ dislocations and voids form times more frequently in case of non-split cascades, which are more localized.
Lastly, according to the analysis of defects and their neighboring atoms, the presence of lighter, smaller elements in WTaCrV is a reason behind a difference in radiation resistance.

\section*{Acknowledgements}
Research was funded in part by the National Science Centre (NCN), Poland, grant number DEC-2023/50/E/ST11/00212/R and the European Union Horizon 2020 research and innovation program under Grant Agreement No. 857470 and from the European Regional Development Fund under the program of the Foundation for Polish Science International Research Agenda PLUS, Grant No. MAB PLUS/2018/8, and the initiative of the Ministry of Science and Higher Education “Support for the activities of Centers of Excellence established in Poland under the Horizon 2020 program” under Agreement No. MEiN/2023/DIR/3795. We gratefully acknowledge Polish high-performance computing infrastructure PLGrid (HPC Center: ACK Cyfronet AGH) for providing computer facilities and support within computational Grant No. PLG/2024/017084, PLG/2025/018326 and PLG/2026/019110. 

\section*{References}
 \bibliographystyle{iopart-num} 
 \bibliography{biblio}

\end{document}